\documentclass[pra,twocolumn,showpacs]{revtex4}
\usepackage{graphicx}
\begin{document}

\bibliographystyle{unsrt}

\title{Experimental Controlled-NOT Logic Gate for Single Photons in the Coincidence Basis }
\author{T.B. Pittman, M.J. Fitch, B.C Jacobs, and J.D. Franson}
\affiliation{Johns Hopkins University,
Applied Physics Laboratory, Laurel, MD 20723}

\date{\today}

\begin{abstract}
We report a proof-of-principle demonstration of a probabilistic controlled-NOT gate for single photons.  Single-photon control and target qubits were mixed with a single ancilla photon in a device constructed using only linear optical elements. The successful operation of the controlled-NOT gate relied on post-selected three-photon interference effects which required the detection of the photons in the output modes.
\end{abstract}

\pacs{03.67.Lx, 42.50.Dv, 42.65.Lm}

\maketitle

There has been considerable interest in a linear optics approach to quantum computing \cite{knill01,franson02}, in which probabilistic two-qubit logic operations are implemented using linear optical elements and measurements made on a set of $n$ additional (ancilla) photons. Here we report a proof-of-principle demonstration of a probabilistic controlled-NOT (CNOT) gate using a single ancilla photon.  Two of the required single-photons were produced using parametric down-conversion \cite{klyshkobook} while a third photon was obtained from an attenuated laser pulse.  The use of only one ancilla photon required that all three photons be detected, in which case the device was known to have correctly performed a CNOT logic operation.

Logic operations are inherently nonlinear, so it is somewhat surprising that they can be performed using simple linear optical elements \cite{knill01,koashi01,ralph01,pittman01,zou01,ralph02,hofmann02,sanaka02}.  The necessary nonlinearity is obtained by mixing the input photons with $n$  ancilla photons using linear elements, and then measuring the state of the ancilla photons after the interaction.  The measurement process is nonlinear \cite{resch01}, since a single-photon detector either records a photon or not, and it projects out the desired logical output state provided that certain results are obtained from the measurements.  The results of the operation are known to be correct whenever these specific measurement results are obtained, which occurs with a failure rate that scales as $\frac{1}{n}$ \cite{knill01} or $\frac{1}{n^2}$ \cite{franson02} in the limit of large $n$, depending on the approach that is used.

In a series of earlier experiments \cite{pittman02b,pittman02a}, we demonstrated several elementary logic gates for single photons, including a quantum parity check and a destructive CNOT gate.  The latter device produced only a single output (the target qubit) and was equivalent to an exclusive OR gate, since the control qubit was destroyed.  As a result, destructive CNOT gates cannot be used for reversible computing or to demonstrate the generation of entanglement, for example.  Here we describe the demonstration of a full CNOT gate whose output includes both the target and control qubits.  Although the gate is probabilistic and the presence of both output qubits must eventually be verified by subsequent measurements (the so-called coincidence basis), the quantum features of both outputs can still be investigated.  For example, the generation of entanglement can be demonstrated using Bell's inequalities and both outputs of such a gate can be fed into subsequent logic gates, provided that one can eventually verify that both outputs were produced. 

\begin{figure}[b]
\includegraphics[angle=-90,width=3in]{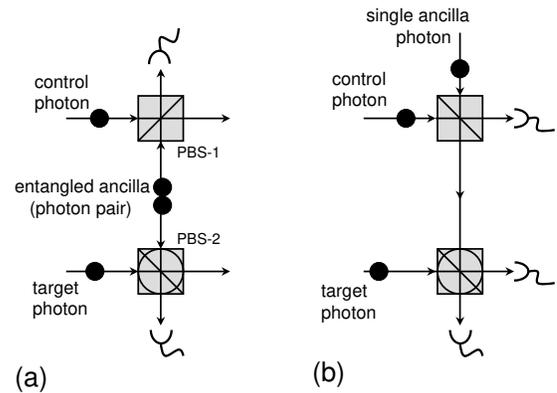}
\vspace*{-.15in}
\caption{Two implementations of a CNOT gate using linear optics and ancilla photons. (a) Our previously proposed gate \protect\cite{pittman01} which relies on two entangled ancilla photons.  (b) The simplified implementation, which requires only one ancilla photon, and is the subject of this paper. PBS-1 and PBS-2 are polarizing beam splitters, with PBS-2 being rotated by $45^{\circ}$.}
\label{fig:overview}
\end{figure}

    We have previously shown \cite{pittman01} that a CNOT gate can be implemented using the simple beam splitter arrangement shown in Figure \ref{fig:overview}(a).  Here the logical value of each of the qubits is represented by the polarization state of a single photon, where a horizontal polarization state $|H\rangle$ represents a value of 0 and a vertical polarization state $|V\rangle$ represents a value of 1.  In addition to the two input photons, a pair of ancilla photons in an entangled state $|\phi^{+}\rangle=\frac{1}{\sqrt{2}}(|00\rangle+|11\rangle)$ are incident on two polarizing beam splitters as shown.  Polarization-sensitive detectors measure the state of the ancilla photons in an appropriate basis when they leave the beam splitters, and corrections to the output may be required based on the results of these measurements \cite{pittman02a}.  Provided that one and only one photon is found in each of these detectors, the output of the device will correspond to that of a CNOT gate \cite{pittman01}.  This corresponds to the case of $n=2$  and the device succeeds in producing the correct output with a probability of $\frac{1}{4}$.

The basic operation of this controlled-NOT gate can be roughly understood as follows:  The lower beam splitter performs the logical function of a controlled-NOT gate except that one of its input qubits is destroyed in the lower detector (a destructive CNOT gate).  The destruction of one of the input qubits can be avoided by first copying (encoding) the value of that qubit onto two output qubits.  One of the copies is then input into the destructive CNOT gate while the remaining copy serves as one of the required logical outputs.  A quantum encoder of that kind can be implemented using the upper beam splitter and the entangled ancilla photons as described in Ref. \cite{pittman01}.  The net result of these operations is a full CNOT gate with both of the input qubits preserved.  

From an experimental point of view, the main difficulty with the CNOT gate of Figure \ref{fig:overview}(a) is its reliance on heralded entangled ancilla pairs  \cite{santori02,stace02,pittman03a}, which have not yet been reliably demonstrated.  However, the need for a pair of ancilla photons in an entangled state can be avoided using the one-ancilla CNOT gate shown in Figure \ref{fig:overview}(b).  This device is equivalent to that of Figure \ref{fig:overview}(a) except that a single ancilla enters the upper beam splitter where a detector was previously located.  If the ancilla photon is in an equal superposition of $|0\rangle$  and $|1\rangle$, then it can be shown that the value of the control photon will be copied (encoded) into the two output ports of the upper beam splitter just as before.  However, the price that must be paid for using a single ancilla photon instead of an entangled ancilla pair is that the correct logical output will only be produced if a single photon actually exits from each of the three output ports of the device. Without reliable quantum non-demolition devices \cite{grangier98}, this condition can only be verified by eventually detecting the output qubit photons as shown Figure \ref{fig:overview}(b).  This type of ``coincidence basis'' operation prohibits the use of the device in a scalable approach to quantum computing, but it does provide a convenient means of demonstrating the operation of a CNOT logic gate for single photons \cite{ralph02}.

The operation of the CNOT gate shown in Figure \ref{fig:overview}(b) can be understood by considering an input state consisting of the single ancilla photon in the required superposition state $\frac{1}{\sqrt{2}}(|0_{A}\rangle + |1_{A}\rangle)$, and an arbitrary initial two-photon state of the control and target photons,  $\alpha_{1}|0_{c}0_{t}\rangle + \alpha_{2}|0_{c}1_{t}\rangle +\alpha_{3}|1_{c}0_{t}\rangle +\alpha_{4}|1_{c}1_{t}\rangle $, where $\sum_{i=1}^{4}|\alpha_{i}|^{2}=1$.  It can be shown that (under ideal experimental conditions) this initial state is transformed into an output state of the form:

\begin{eqnarray}
\lefteqn{|\psi\rangle_{out} =}
\nonumber
\\
& & \frac{1}{2\sqrt{2}}|0_{A}\rangle \left(\alpha_{1}|0_{c}0_{t}\rangle + \alpha_{2}|0_{c}1_{t}\rangle +\alpha_{3}|1_{c}1_{t}\rangle + \alpha_{4}|1_{c}0_{t}\rangle\right)
\nonumber
\\
& &+ \frac{1}{2\sqrt{2}}|1_{A}\rangle 
\left(\alpha_{1}|0_{c}1_{t}\rangle + \alpha_{2}|0_{c}0_{t}\rangle + \alpha_{3}|1_{c}0_{t}\rangle +\alpha_{4}|1_{c}1_{t}\rangle\right) 
\nonumber
\\
& & + \frac{\sqrt{3}}{2}|\psi_{\perp}\rangle 
\label{eq:cnottransform}
\end{eqnarray}

\noindent where $|\psi_{\perp}\rangle$ represents the normalized combination of amplitudes that are orthogonal to the ``coincidence basis'' measurement condition of one photon in each of the three output modes.   

The first term in equation (\ref{eq:cnottransform}) indicates that the detection of a single photon in the $|0\rangle$ state by $D_{A}$ projects the output in the control and target modes into the desired CNOT transform \cite{nielsenchuangbook} of the input. This occurs with a probability of $\frac{1}{8}$, which reflects the probabilistic nature of the device.  The second term in equation (\ref{eq:cnottransform}) shows that the overall success probability could be increased to $\frac{1}{4}$ by also accepting events in which $D_{A}$ registers a single photon in the $|1\rangle$ state, provided that feed-forward control techniques \cite{pittman02a} are used to bit-flip the output target state.

\begin{figure}[b]
\hspace*{-.25in}
\includegraphics[angle=-90,width=3.5in]{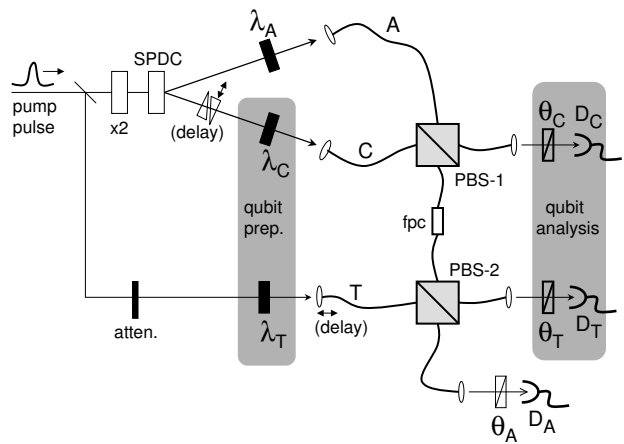}
\vspace*{-.25in}
\caption{Experimental apparatus used to demonstrate the CNOT gate of Fig.\protect\ref{fig:overview}(b). Short laser pulses from a mode-locked Ti:Sapphire laser ($\approx$ 150fs, 76MHz, 780nm) were frequency doubled (x2) to provide UV pulses (390nm) that were used to pump a 0.7mm thick BBO crystal (labelled SPDC) for parametric down-conversion. The down-converted photons were coupled into the ancilla (A) and control qubit (C) single-mode fiber input ports of the upper polarizing beam splitter (PBS-1).  A small fraction of the original pumping pulse was picked off and used as the weak coherent state, which was coupled into the target qubit (T) port of PBS-2. $\lambda_{A,C,T}$ were half-wave plates used for ancilla and qubit state preparation, while $\theta_{A,C,T}$ were polarizers used for post-selection and qubit analysis. $D_{A,C,T}$ were single-photon detectors, which were preceded by 10nm bandpass filters at 780nm (not shown). fpc was a calibrated fiber polarization controller used to rotate the reference frame of PBS-2 by $45^{\circ}$ with respect to PBS-1.}
\label{fig:experiment}
\end{figure}

 A simplified schematic of the experimental apparatus used to demonstrate the CNOT gate of Figure \ref{fig:overview}(b) is shown in Figure \ref{fig:experiment}.  Two of the three photons were produced using a pulsed laser beam passing through a nonlinear crystal (parametric down-conversion \cite{klyshkobook}), while the third photon was obtained by attenuating the laser pulses themselves to the point that each pulse had only a small probability of containing a single photon.  Since the frequency of the laser beam was doubled before the down-conversion process, the down-converted photons had the same frequency as the photons obtained directly from the laser beam.  Furthermore, the use of short laser pulses followed by narrow-band interference filters ensured that all three photons were very nearly indistinguishable \cite{zukowski95,ou97}.

Single-mode optical fibers were used to connect the beam splitters and other components, which reduced the possibility of an error due to mode mismatch.  Errors due to changes in the state of polarization of the photons were minimized using standard optical-fiber polarization controllers.  The required $45^{o}$ rotation of PBS-2 was accomplished by using a calibrated fiber polarization rotator (fpc) between the two beam splitters, and rotating the definitions of $|0\rangle$ and $|1\rangle$ by $45^{o}$ in the remaining ports of PBS-2.  The polarization states of the input photons could be varied by rotating half-wave plates placed before the beams entered the optical fibers, as shown in Figure \ref{fig:experiment}.  This allowed the logical inputs to the device to consist of arbitrary superposition states.

In order for the three photons to be indistinguishable, it was also necessary for them to arrive at the appropriate beam splitters at the same time.  The required path length adjustments were optimized by maximizing the visibilities of various two-photon \cite{hong87,shih88} and three-photon interference effects \cite{rarity97,pittman03b}.  The visibilities of these interference patterns were typically in the range of  85 - 95\%  for two-photon interference effects \cite{hong87,shih88} at PBS-1,  and 60 - 70\%  for three-photon (ie. gated two-photon) interference effects \cite{rarity97,pittman03b} at PBS-2.  The lower values for the three-photon interference effects were primarily due to the use of interference filters with a relatively wide bandwidth of 10 nm \cite{zukowski95,ou97}, and a decreased signal-noise-ratio as described in reference \cite{pittman03b}.  The use of smaller bandwidth filters would be expected to substantially increase the three-photon visibility \cite{pittman03b}, at the cost of lower counting rates.

    The output of the device was measured using polarization analyzers followed by single-photon detectors, and events were only accepted if all three detectors registered a photon. In accordance with Equation (\ref{eq:cnottransform}), $D_{A}$ only accepted ancilla photons in the logical state $|0\rangle$, which was accomplished by fixing the orientation of $\theta_{A}$ at $0^{o}$ in the computational basis (which was physically rotated by $45^{o}$ due to the orientation of PBS-2).  In this initial demonstration, feed-forward control techniques \cite{pittman02a} were not used to accept events in which the ancilla photon was found in the logical state $|1\rangle$. Since the attenuated laser pulses correspond to weak coherent states, it was necessary to minimize the probability of there being two photons in a given pulse by reducing the probability of a single photon 
to roughly $10^{-3}$.

\begin{figure}[t]
\hspace*{-.25in}
\includegraphics[angle=-90,width=4.25in]{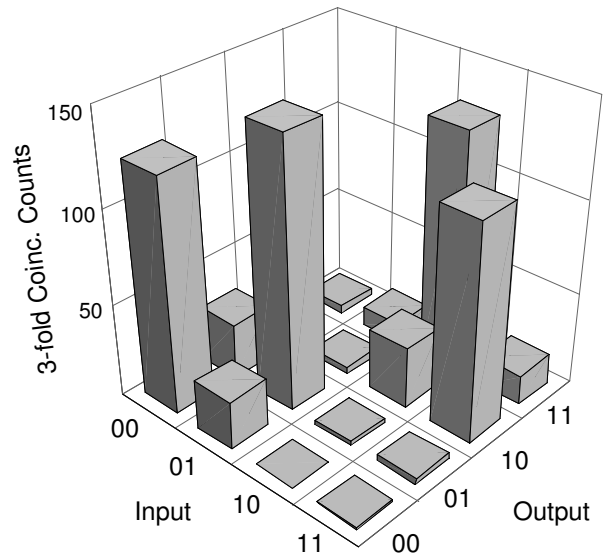}
\vspace*{-.25in}
\caption{Experimental results from the CNOT gate of Figure \protect\ref{fig:overview}(b). The data shows the number of 3-fold coincidence counts per 1500 seconds as a function of the output qubit analyzers for all possible input combination of control and target basis states.}
\label{fig:basisstates}
\end{figure}

Using these techniques, it was possible to measure the output of the device for all possible combinations of logical inputs (0 or 1) as well as superposition states.  The results of these measurements when the input qubits both had specific values of 0 or 1 are summarized in Figure \ref{fig:basisstates}.  The correct results from an ideal CNOT gate \cite{nielsenchuangbook} correspond to the four larger peaks, while the smaller peaks correspond to incorrect results.  It can be seen that the output of the device is clearly correct, aside from overall technical errors of roughly 21\%.  For a given input state, the distribution of errors among the three incorrect output states was primarily determined by the extent to which they depended on destructive three-photon quantum interference effects.  In addition, minor changes in the polarization states of the photons in the optical fibers allowed small contributions from the output state associated with the detection of an ancilla photon in the state $|1\rangle$.   

\begin{figure}[t]
\hspace*{-.1in}
\vspace*{-.5in}
\includegraphics[angle=-90,width=3.25in]{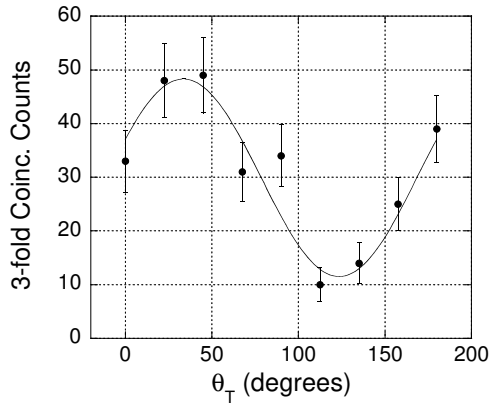}
\vspace*{.3in}
\caption{Typical experimental results obtained using a superposition state for the control qubit, which is expected to produce an entangled output state, $\frac{1}{\sqrt{2}}(|00\rangle+|11\rangle)$. The data shows the number of three-fold coincidence counts per 1500 seconds as a function of the target analyzer, with the control analyzer fixed at $0^{\circ}$ (qubit value $|0\rangle$). The solid line is a sinusoidal least squares fit to the data, with a visibility of (61.5$\pm$7.4)\%. The slight shift away from the expected target value of $|0\rangle$ (eg. $\theta_{T}=45^{\circ}$) was primarily due to incompletely compensated birefringences in the fibers.}
\label{fig:coherence}
\end{figure}

    It is important to demonstrate that quantum logic gates maintain the quantum-mechanical coherence of the input qubits when the latter are in superposition states of 0 and 1.  As an example of this coherence, Figure \ref{fig:coherence} shows the results obtained when the incident control qubit was in the superposition state $\frac{1}{\sqrt{2}}(|0\rangle+|1\rangle)$ while the target was in the state $|0\rangle$.  In that case, the two output qubits should be produced in an entangled state of the form $\frac{1}{\sqrt{2}}(|00\rangle+|11\rangle)$, which is the state $|\phi^{+}\rangle$.  This is an important example, since a CNOT gate is expected to produce entanglement between two independent input photons.  The data in the figure corresponds to the number of three-fold coincidence counts as a function of the target analyzer, with the control analyzer set to the logical value 0. In that case, a detection of a control photon collapses the entangled state to just the first term, so that the target photon should also be found with logical value 0.  The data of Figure \ref{fig:coherence} are consistent with that prediction.  When the analyzer in the path of the control photon was set to logical value 1 instead, then the entangled state collapsed to the second term and the target photon was found with logical value 1, as expected. Analogous results were found in a basis rotated by $45^{\circ}$. Although these results demonstrate entanglement between the two photons and nonlocal measurement results, the data extracted from these plots was not sufficient to allow a violation of Bell's inequality using the two output photons \cite{clauser78}.

    In summary, we have experimentally demonstrated the operation of a probabilistic CNOT gate for single photons in the coincidence basis using linear optical elements and a single ancilla photon.  Although the presence of the output qubits must eventually be verified, the properties of both the control and target qubits can be investigated before they are detected.  Input qubits with specific values in the computational basis produced output states that corresponded to those of a CNOT gate aside from technical errors on the order of 20\%.  Superpositions of input states were found to produce entanglement of the output qubits.  Although the use of a single ancilla photon limits the performance of the current device, which is not scalable, larger numbers of ancilla photons can be combined with more general linear optics techniques \cite{knill01,franson02} to reduce the probability of error and produce an approach that is scalable.

This work was supported by ARO, NSA, ARDA, ONR, and IR\&D funding.



\end{document}